\documentclass{aa}  
\usepackage{graphicx}
\usepackage{graphbox} 
\usepackage{eqnarray,amsmath}
\usepackage{mathtools}
\usepackage{multicol}
\usepackage{soul}
\usepackage[hypcap]{caption}
\usepackage{txfonts}
\usepackage{subcaption}
\usepackage{arydshln}
\usepackage{hyperref}
\usepackage[svgnames]{xcolor}
\usepackage{ulem}
\usepackage{cancel}
\usepackage[export]{adjustbox}
\graphicspath{{./figures/}} 

\def\msol{\ensuremath{M_\odot}}
\begin{document} 
\defcitealias{vink01}{V01}
\defcitealias{alex23}{GM23}
\defcitealias{krticka24}{K24}
\defcitealias{dejager88}{dJ88}
\defcitealias{yang23}{Y23}

\titlerunning{Revisiting the Evolutionary Status of Massive Stars at the central parsec of the Milky Way}
\title{Revisiting the Evolutionary Status of Massive Stars at the central parsec of the Milky Way}

\authorrunning{Gormaz-Matamala et al.}
\author{A. C. Gormaz-Matamala\inst{1}
\and
J. Cuadra\inst{2,4}
\and
B. Kubátová\inst{1}
\and
J. Kubát\inst{1}
\and
S. Ekström\inst{3}}
\institute{Astronomický ústav, Akademie věd Ceské republiky, Fričova 298, 251 65 Ondřejov, Czech Republic\\
\email{alex.gormaz@asu.cas.cz}
\and
Departamento de Ciencias, Facultad de Artes Liberales, Universidad Adolfo Ib\'a\~nez, Av. Padre Hurtado 750, Vi\~na del Mar, Chile
\and
Department of Astronomy, University of Geneva, Chemin Pegasi 51, 1290 Versoix, Switzerland
\and
Millennium Nucleus on Transversal Research and Technology to Explore Supermassive Black Holes (TITANS), Chile
}

\date{}

\abstract 
{Massive stars and their winds have a large influence in their environment. For example, they determine the accretion rate on to the Galactic Centre (GC) super-massive black hole Sagittarius A* (Sgr A*).
The winds of those stars collide and are accreted, at a rate that depends on their chemical composition.
Also, the new self-consistent approach to modelling stellar winds of these stars leads to lower mass-loss rates compared to previous standard values, thus altering the stellar properties of their advanced evolutionary stages.}
{Here we aim to revisit the evolutionary status of the evolved massive stars at the GC, by means of new tracks based on updated mass-loss rate recipes for the earlier stages of massive stars.}
{We use the Geneva-evolution-code (\textsc{Genec}) for initial stellar masses ranging from 20 to 60 $M_\odot$, for metallicity $Z=0.020$.
We adopt a new mass-loss rate recipe for the line-driven winds of O-type stars and B-supergiants, plus a new recipe for the dust-driven winds of red supergiants (RSG).
Additionally, we set up initial rotation $\Omega/\Omega_\text{crit}=0.4$, and we adopt Ledoux criterion for the treatment of convection in inner layers.}
{We found that evolution models adopting new mass-loss rate prescriptions predict that stars will lose less of their outer layers during their initial phases, while a big reduction of mass happens at the RSG phase.
As a consequence, the resulting Wolf-Rayet (WR) stars are less radially homogeneous in their inner structure from the core to the surface.
Also, these new evolution models predict the absence of hydrogen-free WN stars.
These evolutionary predictions agree better with the observed properties of the WR stars at the GC, in special with their chemical abundances.}
{We provide a table with the chemical H, He, and CNO abundances calculated for the different subtypes of WR stars (Ofpe/WN9, WNL, WN/C, and WC).
We propose a different re-arrangement of the WR subtypes to be used for the modelling of the collision of their winds.
We discuss the potential implications of these changes for the colliding winds generated from the massive stars at the GC, which are accreting onto the supermassive black hole Sgr A*.}


\maketitle

\section{Introduction}	
	Massive stars, born with $M_\text{zams}\ge8\,M_\odot$, are the hottest and most luminous stars in the Universe. 	They lose much of their initial mass due to stellar winds and outflows, which are crucial not only for determining the star’s final mass, but also for influencing their evolutionary path.
	These stars are crucial for studying nucleosynthesis, the production of ionising flux, feedback from wind momentum, star formation history, and galaxy evolution.

	Over the past decade, considerable effort has gone into developing new theoretical mass-loss rate recipes based on hydrodynamic wind calculations \citep{kk17,alex19,bjorklund21}, with the aim of updating stellar evolutionary models of massive stars. 
	The mass-loss rates (denoted $dM/dt$ or simply $\dot M$) of these new prescriptions are a factor $\sim2-3$ times lower than the smooth rates given by \citet[hereafter \citetalias{vink01}]{vink01}, which have been the standard recipe adopted in the majority of stellar evolution codes such as \textsc{Genec} \citep[e.g.,][]{eggenberger08}, \texttt{MESA} \citep[e.g.,][]{paxton11}, or BoOST \citep[e.g.,][]{szecsi22}.
	Evolution models adopting lower $\dot M$ show that massive stars retain more mass during their main sequence phase, thus being bigger and more luminous \citep{alex22b,bjorklund23}.
	Moreover, stars keep larger rotational velocities for a more extended time because they lose less angular momentum \citep{alex23,alex24a}, also altering the chemical enrichment at the stellar surface due to rotational mixing.
	The impact of reduction in $\dot M$ goes from different chemical abundances \citep{vmagg24,liu25}, to larger final masses prior to the core-collapse \citep{alex24b,romagnolo24,kruckow24,costa25}.
	Even though these changes in the stellar evolution have been applied mostly at the early H-burning stages, they also impact the evolution of the subsequent stages \citep{josiek24}, notwithstanding the importance of upgrading mass-loss rate prescriptions for these advanced stages such as red supergiants \citep{zapartas24} or Wolf-Rayet (WR) stars \citep{sander23}.

	New models for stellar winds and evolution modify the diagnostics for populations of massive stars.
	A prime example is the central parsec of the Galactic Centre (hereafter GC), with the largest concentration of massive stars known in the Milky Way.
	\citet{bartko09} identified spectroscopically 90 O/WR stars, while \citet{vonfellenberg22} increased the sample to 195 but including young B stars also.
	At the same time, this population is puzzling because, even though they are evolved, the stars are `young' (aged $\sim2-8$ Myr) but living in a region hostile for star formation \citep[e.g.,][]{ghez03,genzel10}, given the tidal field of a supermassive black hole, Sgr~A*, with a mass $M_\text{BH}\approx 4\times10^{6}\,M_\odot$ \citep{schodel02}.
	Moreover, this population has a very unusual distribution of luminosities (after correcting for extinction), which implies a very top-heavy initial mass function \citep[e.g.,][]{paumard06, nayakshin06, bartko10, lu13, vonfellenberg22}.
	Therefore, it is worth revising the observational constraints on the ages and initial masses.

	The GC also provides the best chance to study a galactic nucleus, including the accretion on to a super-massive black hole, which in this case is closely connected to the massive stars and their winds.
	The stars are all confined to a fraction of a parsec, so the winds are expected to collide, convert their kinetic energy into thermal, and create a hot plasma from which the black hole accretes \citep{quataert04}.
	Different groups have modelled the gas dynamics in the GC, using the properties of each star, including their orbits and wind characteristics, as initial conditions \citep[e.g.,][]{cuadra08, cuadra15, ressler18, ressler20, calderon20apj, calderon25}.
	These studies conclude that the accretion onto Sgr A* is more complicated and time-variable, and depends on the wind properties.
	In particular, of the Wolf-Rayet winds at the central parsec, given that they possess much stronger winds than OB-type stars.
	For the WR winds with terminal velocity $\gtrsim1000$ km s$^{-1}$, when they collide gas is heated to $\gtrsim10^7$ K and cooling takes much longer than the dynamical time, leading to a smooth, Bondi-like accretion.
	Conversely, after the collision of winds with $\varv_\infty\lesssim1000$ km s$^{-1}$, gas is heated to $\sim10^6$ K only, and the cooling process is faster, leading to the formation of clumps of mass up to the order of $10^{-5}\,M_\odot$ \citep{cuadra05,calderon16,calderon20mn}, some of which may have been observed \citep{gillessen12, gillessen25}.
	Since clumps are present, there are episodes of high accretion superimposed on the smooth accretion rate of the hot plasma.
	If many clumps reach the black hole vicinity, they can coalesce forming a cold disk-like structure around the black hole Sgr A*, making its growth more efficient \citep{calderon20apj,calderon25}.
	The chemical composition of the winds is also an important factor.
	WR stars are losing their atmospheric hydrogen, so their winds can have a hydrogen fraction as high as $\approx50\%$ and as low as zero.
	Hydrogen significantly influences the cooling of the gas after the winds shock, as \cite{calderon25} demonstrated, potentially tipping the balance on whether a disc forms around the black hole.
        Observationally, studies using different tracers have found evidence for and against the current presence of such a cold disc around Sgr~A* \citep{Murchikova19, Ciurlo21}.

	While there have been many studies of the dynamics and origin of the young stellar population in the GC \citep[see, e.g., the review by][]{mapelli16}, their atmospheric properties have not received that much attention.
	The properties of the B-type stars at the GC were studied by \citet{habibi17}, whereas the O and WR stars were analysed by \citet{martins07}.
	In the latter, the evolutionary sequence for the different subtypes of WR stars (Ofpe/WN9, WNL, WNC, WC) was analysed with the evolutionary tracks from \citet{meynet05}, where the outflows for the massive OB-type stars are overestimated.
	Therefore, the calculation of new self-consistent evolutionary tracks, including a new prescription for $\dot M$, is necessary to further our understanding of not only massive stars and their stellar winds, but also their effects over the interstellar medium and the current growth of the Galactic central black hole.

	In this paper, we present a new set of evolutionary tracks adopting new mass-loss rate recipes for the initial OB-type and RSG phases.
	We describe the physics of our evolution models in Sec.~\ref{evolutionmodels} and analyse their results in Sec.~\ref{results}.
	We compare our predictions with the observed properties of the stars at the GC in Sec.~\ref{stellarproperties}, together with provided insights of the chemical composition and lifetimes of the calculated WR phases.
	Finally, a summary and conclusions are outlined in Sec.~\ref{conclusions}.

\section{Evolution models}\label{evolutionmodels}
	To approach the chemical conditions close to the GC, we model the stellar evolution of stars at super-solar metallicity $Z=0.020$ \citep[being $Z_\odot=0.0142$ from][]{asplund09}, rescaling $Z$ as done by \citet{yusof22}.
	The metallicity of the Milky Way is not constant but it decreases according to the Galactocentric distance, with $Z=0.028\pm0.004$ near the GC \citep{hayden14}.
        Measurements at the GC itself are not conclusive, but seem to indicate a $\sim 50\%$ suprasolar abundance of alpha elements and a solar iron abundance \citep[see][for a review]{genzel10}.
	Our selected $Z=0.020$ is perhaps a simplification; however, it enables us to make a direct comparison with earlier grids of evolution models such as \citet{yusof22}, which is the grid with the highest metallicity currently in the literature.

\subsection{Mass-loss rate prescription}
	For OB-type stars we use the mass-loss rate recipe from \citet[hereafter \citetalias{krticka24}]{krticka24}.
	This recipe comes from models calculated for O-type stars using the \textsc{Metuje} wind modelling code \citep{kk17,kk18}, and shows a close agreement with the mass-loss rate coming from \citet[hereafter \citetalias{alex23}]{alex23} for main sequence stars.
	However, the formula from \citetalias{alex23} is valid for O-type stars with $T_\text{eff}\ge30$ kK only, whereas \citetalias{krticka24}'s is valid for OB-type stars with $T_\text{eff}>10$ kK, thus covering also the range of B-supergiants \citep{krticka21}.
	The so-called bi-stability jump \citep{pauldrach90} is expected to take place within this temperature range, even though its existence has been debated lately.
	\citet{vink01} predict an abrupt increase in mass-loss rate at $T_\text{eff}\simeq 25$~kK due to the change in the wind ionisation \citep{vink99}, whereas \citet{bjorklund23} predict just a decrease in the mass-loss rate with decreasing temperature without jumps.
	On its side, the mass-loss rate prescription from \citet{krticka24} finds a gradual rise in $\dot M$ also due to the change in the Fe ionisation, below $22$ kK with a peak at $T_\text{eff}\simeq15$ kK, thus more in rule with the observational surveys which have found no drastic jumps in the mass-loss rate of OB-supergiants either in Galactic sources \citep{deburgos24} or in the LMC \citep{verhamme24}.

	It is worth to mention that \textsc{Metuje} models have been calculated in the metallicity range from $0.2\,Z_\odot$ to $1\,Z_\odot$, and thus the \citetalias{krticka24} mass-loss rate recipe used in this work is an extrapolation towards super-solar metallicity.
	Standard evolution models adopting $\dot M_\text{V01}$ assume a fixed metallicity dependence of $\dot M\propto Z^{0.85}$ for OB-type stars; however, new prescriptions demonstrate that such metallicity dependence varies implicitly with the luminosity or the stellar mass \citep{kk18,alex23}, given that line driving (metallicity dependent) becomes less dominant than electron scattering (metallicity independent) at larger luminosities, something also observed for WNL stars \citep{grafener08}.
	For the purposes of this paper, given that the increase in metallicity is only a $\approx42\%$ from 0.014 to 0.020, the extrapolation adopted by us is still within the expected order of magnitude for the weaker winds.

	We also update the mass-loss rate recipe for the region of the Hertzsprung-Russell diagram (HRD) below 10 kK. 
	We keep the default formulation from \citet[hereafter \citetalias{dejager88}]{dejager88} for yellow supergiants ($5 < T_\text{eff} < 10$ kK, hereafter YSG), but adopt a new recipe for red supergiants ($T_\text{eff} < 5$ kK, hereafter RSG).
	Unlike line-driven winds, the mechanism for the mass loss of RSG is still uncertain, thought to be a product of dust-driven and low surface gravity \citep{vanloon25}.
	Consequently, predictions of $\dot M_\text{RSG}$ show discrepancies of various orders of magnitude \citep{goldman17,yang23,beasor23,decin24}, as well as a strong dependence on the assumed turbulent velocity \citep{kee21} or grain size for dust-driven simulations \citep{antoniadis24}, thus leading into different evolutionary outcomes.
	Higher $\dot M_\text{RSG}$, such as \citet{kee21} and \citet{yang23}, predict that the stars will remove their outer envelopes faster resulting in a moderate population of RSG more in agreement with observations\footnote{Notice that `higher mass-loss rate' means `higher mass-loss rate at the same HRD location'. The total RSG mass loss may change if the stellar track reaches the red region of the HRD at lower luminosities, as we will see later.} \citep{zapartas24}.
	Despite these scattered results, we incorporate the mass-loss rate for red supergiants ($T_\text{eff} < 5$ kK) from \citet[hereafter \citetalias{yang23}]{yang23} in our study, as it aligns with the updated Humphreys-Davidson limit for the SMC \citep{davies18} and M31 \citep{mcdonald22}.
	Study of \citet{yang23} is also focused on the SMC, but since $\dot M_\text{RSG}$ is expected to be independent of $Z$ \citep{antoniadis25} the formula can be easily extended for other metallicities.
	Besides, a higher mass-loss rate prescription for RSG intrinsically incorporates the existence of sporadic gaseous ejections, which can reach $\dot M\sim3\times10^{-3}$ $M_\odot$ yr$^{-1}$ in some extreme cases \citep{humphreys22}.

	For WR winds we keep the mass-loss rate recipes from \citet{nugis00} and \citet{grafener08}.
	The reason is that, for the mass range to be studied ($M_\text{zams}$ from 20 to 60 $M_\odot$) the WR stars are post RSG, with the Ofpe/WN9 spectral type \citep{crowther95} as an intermediate phase.
	More recent formulae for $\dot M$ of the WR stars have been explored for the cases where the transition between O-type and WR stages is given by the WNh phase \citep{bestenlehner20} or for H-depleted WR stars \citep{sander20b}.
	Even though these new recipes have been included in studies about stellar evolution \citep{romagnolo24}, more detailed analyses on the expected lifetimes and chemical composition of the resulting WR stars have been focused only on very massive stars \citep{alex24b}.
    
	The formulae described in this Section are given explicitly in Appendix~\ref{mdot_formulae}.

\subsection{Additional setup}
	We calculate our evolution models using the Geneva-evolution-code \citep[hereafter \textsc{Genec}]{eggenberger08}.
	We keep most of the setup for the inner stellar structure from \citet{ekstrom12} and \citet{yusof22}, such as step-overshooting $\alpha_\text{ov}=l/H_p=0.1$, whereas we employ the Ledoux criterion for the convective boundaries, in agreement with \citet{georgy14}, \citet{martinet23}, and \citet{sibony23}.
	
	For the treatment of stellar rotation, we keep the original setup from \citet{ekstrom12}, with the angular momentum transport described by the advective equations from \citet{zahn92},
	Horizontal and vertical shear diffusion coefficients following \citet{zahn92} and \citet{maeder97} respectively, notwithstanding that different diffusion treatments would lead into diverse results \citep{nandal24}.
	The enhancement of mass-loss due to rotation is given by the correction factor from \citet{maeder00}, which is of the same order as the mass-loss increase described by analytical solutions from the m-CAK theory \citep{michel04,michel07}.
	The standard initial rotation for the different masses is set to be a $40\%$ of the critical rotational velocity \citep{ekstrom12,choi16}.
	However, since \textsc{Genec} incorporate the oblate effects in the stellar structure \citep{georgy11}, $\varv/\varv_\text{crit}\neq\Omega/\Omega_\text{crit}$ and thus the initial $\varv/\varv_\text{crit}=0.4$ from \citet{yusof22} actually is equivalent to $\Omega/\Omega_\text{crit}\simeq0.58$.
	Hence, we set the initial rotational velocity for our models to be $\Omega/\Omega_\text{crit}=0.4$ ($\varv/\varv_\text{crit}\simeq0.28$), matching with other evolution grids such as \citet{choi16} and \citet{romagnolo24}.

\section{Results}\label{results}
    \begin{figure}[t!]
		\centering
		\includegraphics[width=1.0\linewidth]{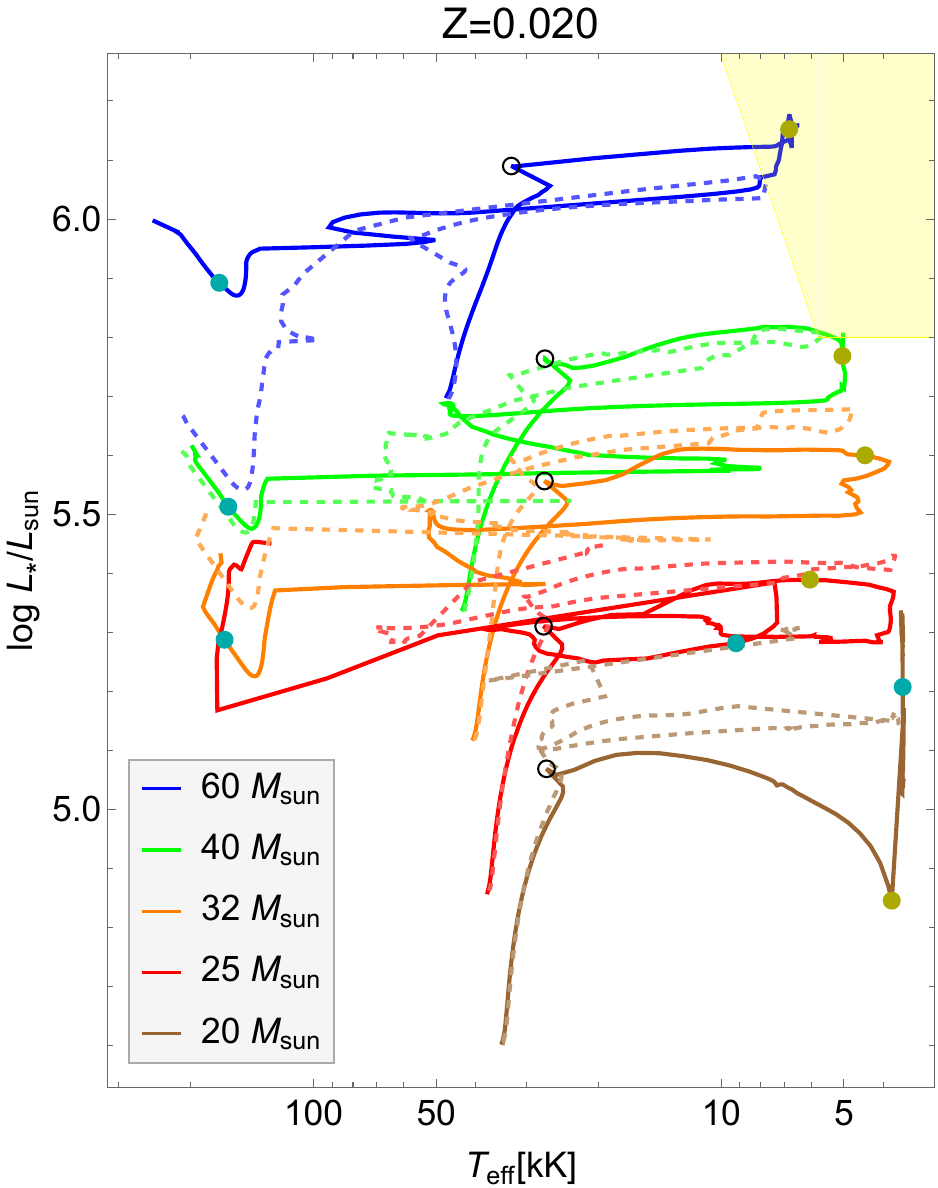}
		\caption{\small{HR diagram for the new evolution models of this work (solid lines), compared with the old models from \citet[][dashed lines]{yusof22}.
		Empty, dark yellow, and dark cyan circles represent the end of the H-core, the beginning of the He-core, and the end of the He-core burning stages, respectively.
		The yellow-shaded region corresponds to the zone beyond the Humphreys-Davidson limit, where LBV stars are expected to be found \citep{humphreys16}.}}
		\label{HRD_z20}
        \end{figure}
	\begin{table*}[t!]
		\centering
		\caption{Properties of our selected evolutionary tracks at the end of the H-core burning stage.}
		\begin{tabular}{ccc|ccccccccccc}
			\hline
			\hline
			$M_\text{zams}$ & & $\log\dot M_\text{zams}$ & $\varv_\text{rot,ini}$ & $\langle\varv_\text{rot}\rangle$ & $\varv_\text{rot,final}$ & $\tau_\text{MS}$ & $M_\text{total}$ & $X_\text{H}$ & $X_\text{He}$ & $X_\text{C}$ & $X_\text{N}$ & $X_\text{O}$\\
			$[M_\odot]$ & & & \multicolumn{3}{c}{[km s$^{-1}$]} & [Myr] & $[M_\odot]$ & \multicolumn{5}{c}{mass percentage}\\
			\hline
			$60$ & Y22 & $-5.60$ & 338.0 & 120.3 & 5.2 & 4.367 & 34.86 & 22.39 & 75.65 & 0.014 & 1.152 & 0.027\\
			$60$ & this work & $-6.29$ & 234.0 & 166.6 & 3.5 & 4.027 & 55.22 & 49.33 & 48.69 & 0.127 & 0.825 & 0.336\\
			\\
			$40$ & Y22  & $-6.16$ & 302.0 & 138.6 & 5.0 & 5.600 & 30.59 & 44.52 & 53.51 & 0.069 & 0.939 & 0.202\\
			$40$ & this work  & $-6.72$ & 215.0 & 176.2 & 69.5 & 5.089 & 38.24 & 61.62 & 36.39 & 0.213 & 0.521 & 0.569\\
			\\
			$32$ & Y22  & $-6.52$ & 297.0 & 162.6 & 9.0 & 6.750 & 26.90 & 50.24 & 47.79 & 0.112 & 0.776 & 0.331\\
			$32$ & this work  & $-7.02$ & 205.0 & 173.7 & 130.0 & 5.952 & 31.00 & 66.68 & 31.32 & 0.254 & 0.365 & 0.694\\
			\\
			$25$ & Y22  & $-6.99$ & 277.0 & 191.6 & 35.4 & 8.119 & 23.12 & 60.49 & 37.52 & 0.163 & 0.543 & 0.530\\
			$25$ & this work  & $-7.40$ & 195.0 & 164.6 & 160.0 & 7.208 & 24.51 & 68.92 & 29.07 & 0.270 & 0.283 & 0.765\\
			\\
			$20$ & Y22  & $-7.48$ & 267.0 & 209.2 & 133.0 & 9.602 & 19.37 & 67.06 & 30.93 & 0.193 & 0.381 & 0.676\\
			$20$ & this work  & $-7.80$ & 185.0 & 154.9 & 167.0 & 9.018 & 19.76 & 69.78 & 28.20 & 0.272 & 0.251 & 0.799\\
			\hline
		\end{tabular}
		\label{table_MS_z20}
		\tablefoot{Values for tracks adopting Vink's formula are taken from \citet{yusof22}.}
	\end{table*}
	\begin{table*}[t!]
		\centering
		\caption{Properties of our selected evolutionary tracks at the end of the He-core and the C-core burning stages.}
               \resizebox{\linewidth}{!}{
		\begin{tabular}{cc|ccccccc|ccccccc}
			\hline
			\hline
			$M_\text{zams}$ & $\dot M$ recipe & Age & $M_\text{total}$ & $X_\text{H}$ & $X_\text{He}$ & $X_\text{C}$ & $X_\text{N}$ & $X_\text{O}$ & Age & $M_\text{total}$ & $X_\text{H}$ & $X_\text{He}$ & $X_\text{C}$ & $X_\text{N}$ & $X_\text{O}$\\
			$[M_\odot]$ & & [Myr] & $[M_\odot]$ & \multicolumn{5}{c}{mass percentage} & [Myr] & $[M_\odot]$ & \multicolumn{5}{c}{mass percentage}\\
			\hline
			$60$ & Y22 & 4.768 & 12.89 & 0 & 26.78 & 50.31 & 0 & 20.33 & 4.769 & 12.88 & 0 & 26.65 & 50.32 & 0 & 20.44\\
			$60$ & this work & 4.394 & 21.39 & 0 & 24.00 & 46.11 & 0.001 & 27.24 & 4.398 & 21.30 & 0 & 23.62 & 46.02 & 0.001 & 27.71\\
			\\
			$40$ & Y22  & 6.037 & 11.65 & 0 & 27.16 & 51.08 & 0 & 19.17 & 6.039 & 11.63 & 0 & 27.07 & 51.09 & 0 & 19.25\\
			$40$ & this work  & 5.540 & 11.86 & 0 & 24.82 & 48.94 & 0.026 & 23.61 & 5.546 & 11.78 & 0 & 24.31 & 48.95 & 0.024 & 24.11\\
			\\
			$32$ & Y22  & 7.271 & 9.83 & 0 & 25.82 & 50.48 & 0.033 & 21.07 & 7.274 & 9.80 & 0 & 25.51 & 50.55 & 0.030 & 21.32\\
			$32$ & this work  & 6.493 & 8.49 & 0 & 27.03 & 49.02 & 0.106 & 21.26 & 6.501 & 8.41 & 0 & 26.31 & 49.12 & 0.098 & 21.88\\
			\\
			$25$ & Y22  & 8.736 & 9.16 & 0 & 95.71 & 0.888 & 2.336 & 0.166 & 8.739 & 9.08 & 0 & 95.63 & 1.047 & 2.244 & 0.175\\
			$25$ & this work  & 7.940 & 8.86 & 0.003 & 71.91 & 9.973 & 7.771 & 8.907 & 7.950 & 8.68 & 0 & 38.42 & 24.44 & 0.502 & 33.49\\
			\\
			$20$ & Y22  & 10.425 & 7.35 & 31.54 & 66.49 & 0.009 & 1.014 & 0.198 & 10.435 & 7.27 & 21.95 & 76.08 & 0.011 & 1.100 & 0.101\\
			$20$ & this work  & 9.799 & 12.26 & 60.12 & 37.88 & 0.160 & 0.555 & 0.602 & 9.812 & 11.81 & 59.96 & 38.04 & 0.157 & 0.560 & 0.599\\
			\hline
		\end{tabular}}
		\label{table_HeC_z20}
		\tablefoot{Values for tracks adopting Vink's formula are taken from \citet{yusof22}.}
	\end{table*}
	\begin{figure*}[t!]
		\centering
		\includegraphics[height=6cm,valign=c]{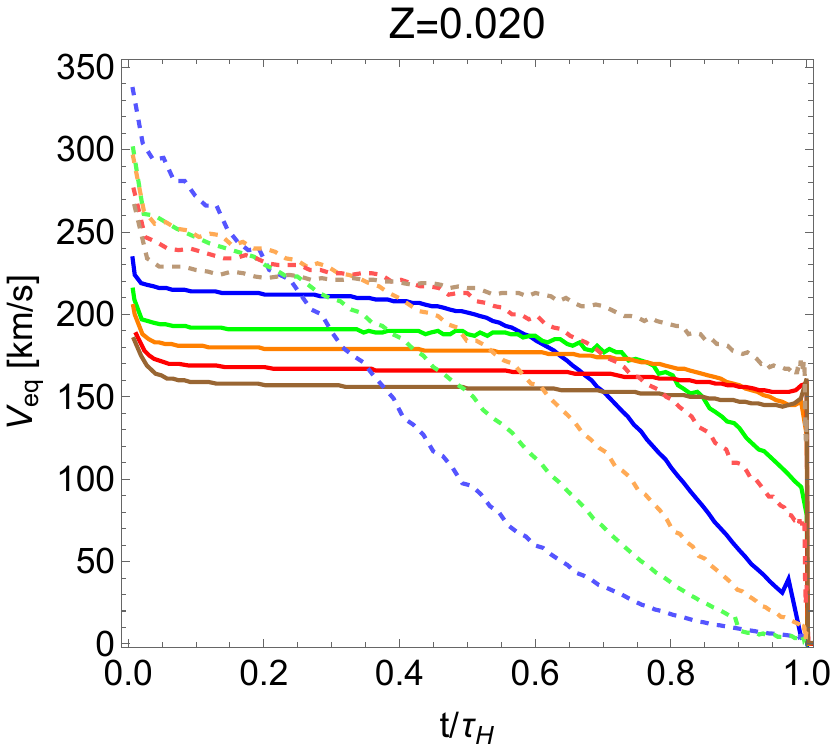}
		\hspace{8mm}
		\includegraphics[height=6cm,valign=c]{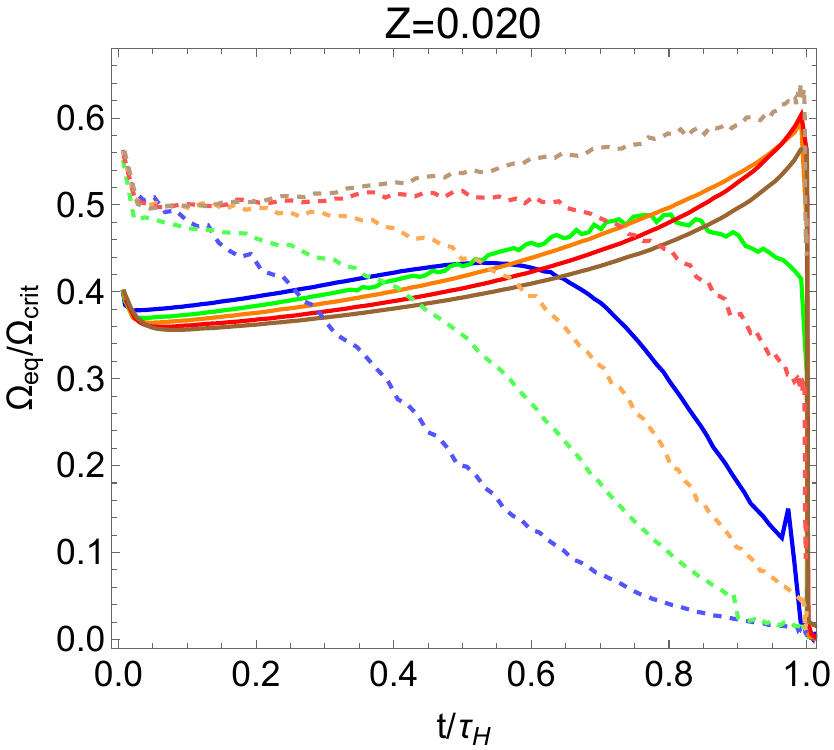}
		\hspace{4mm}
		\includegraphics[width=0.1\linewidth,valign=c]{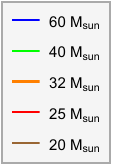}
		\caption{\small{Evolution of rotational velocity expressed as absolute magnitude $\varv_\text{rot}$ (left panel) and angular velocity as fraction of the critical velocity $\Omega/\Omega_\text{crit}$ (right panel), as a function of the H-core burning lifetime $\tau_\text{H}$.
		Solid and dashed lines represent new and old models respectively, same as for Fig.~\ref{HRD_z20}.}}
		\label{rotation_z20_fin}
	\end{figure*}
	\begin{figure*}[t!]
		\centering
		\includegraphics[height=4.5cm,valign=c]{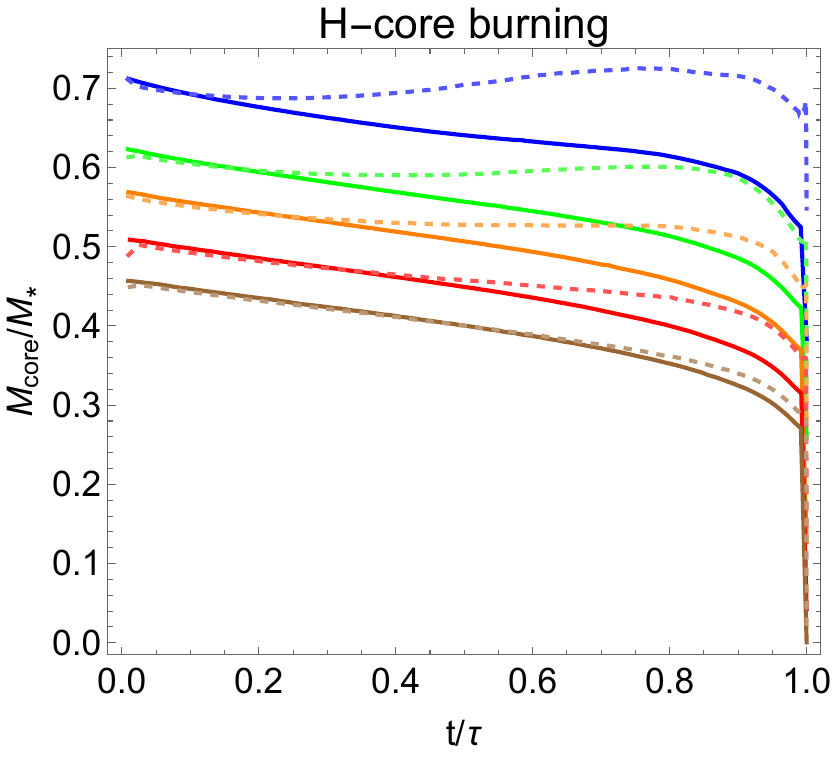}
		\hspace{2mm}
		\includegraphics[height=4.5cm,valign=c]{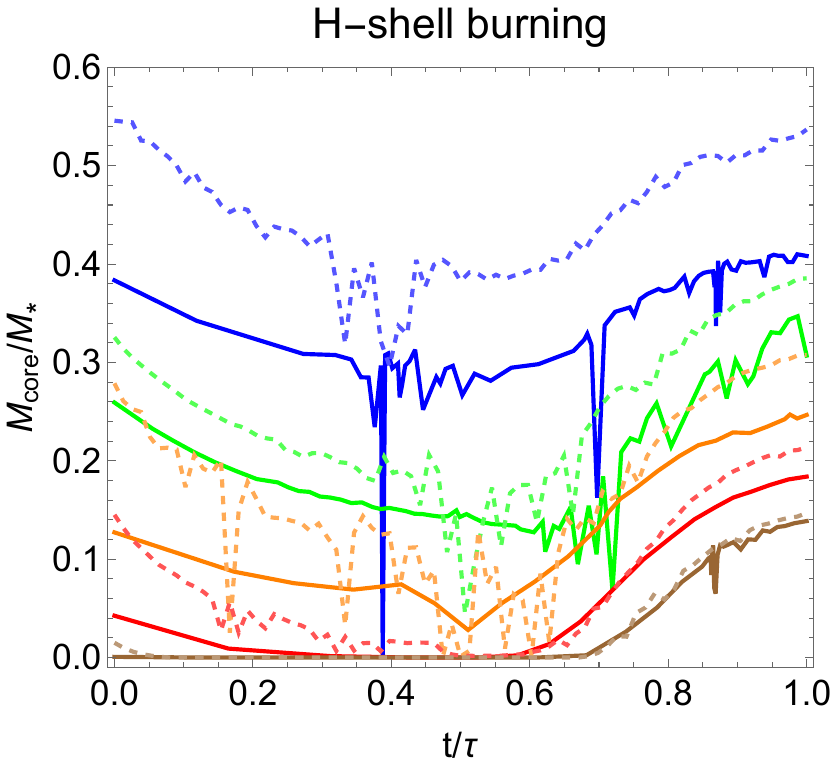}
		\hspace{2mm}
		\includegraphics[height=4.5cm,valign=c]{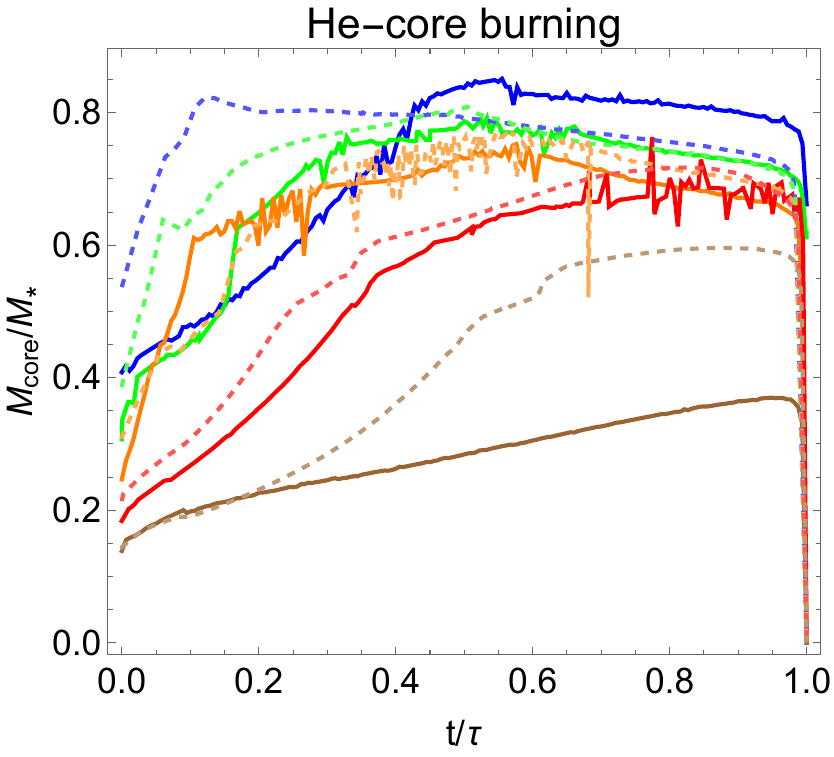}
		\hspace{1mm}
		\includegraphics[width=0.075\linewidth,valign=c]{stellarM_legends}
		\caption{\small{Evolution of the convective cores, as fractions of the total stellar mass, for our models during the H-core, H-shell and He-core stages.
		Solid and dashed lines represent new and old models respectively, same as for Fig.~\ref{HRD_z20}.}}
		\label{coremass_z20}
	\end{figure*}
	Figure~\ref{HRD_z20} shows our new evolution tracks across the HR diagram, together with the tracks from \citet{yusof22}, who used $\dot M_\text{V01}$ for the winds of OB-type stars.
	The main results from both old and new evolution models are tabulated in Table~\ref{table_MS_z20} for the end of the H-core burning, and in Table~\ref{table_HeC_z20} for the end of He-core and C-core burning stages.
	The evolution of the stellar rotation during the H-core burning stage, in absolute terms and relative to the critical rotational velocity, is shown in Fig.~\ref{rotation_z20_fin}.
	The mass of the convective cores as fractions of the stellar masses during the H-core, H-shell, and He-core burning stages, are shown in Fig.~\ref{coremass_z20}.

\subsection{Evolution during the main sequence stage}
	In line with \citet{kk18} and \citet{alex22a}, we see that the mass-loss rates from \citetalias{krticka24} are $\sim3$ times (for $M_\text{zams}=20\,M_\odot$) to $\sim5$ times (for $M_\text{zams}=60\,M_\odot$) weaker than \citetalias{vink01}.
	For main sequence stars, as also observed for low metallicity environments \citep{alex24a}, a reduction of mass loss also implies a reduction in the loss of angular momentum, resulting in a nearly constant rotational velocity during the first half of the main sequence prior to the final decrease in velocity, contrary to the abrupt initial deceleration predicted by older models shown in Fig.~\ref{rotation_z20_fin}.
	This is the reason why, despite starting our models with lower $\varv_\text{rot,ini}$ in comparison with \citet{yusof22}, our models have a larger average rotational velocity $\langle\varv_\text{rot}\rangle$ for the most massive cases.
	
	The combination of both conditions, lower initial $\dot M$ and lower initial $\varv_\text{rot}$, produces a less homogeneous inner structure.
	Reduced mass loss makes stars to retain more of their outer envelopes, whereas lower rotation produces less efficient internal mixing.
	Thus, the models show redder tracks through the main sequence as seen in Fig.~\ref{HRD_z20} (and thus a larger radial expansion) as a consequence of the less homogeneous chemical evolution \citep{maeder87c,meynet13}, thus ending the main sequence with cooler stellar temperatures.
	This scenario was also observed when only isolated changes in $\dot M$ formula were implemented \citep{alex23}, demonstrating that the lower mass loss is far more influential than the reduced rotation.
	A less efficient mixing also means a lower lifetime for the H-core burning process, because the core is less likely to be refilled with hydrogen from the upper layers, and also a more moderate chemical enrichment in the stellar surface, where the new models exhibit larger fractions of hydrogen in comparison with the models from \citet{yusof22}.
	Similarly, the CNO abundances at the stellar surface are affected.
	The H-burning CNO bi-cycle causes nitrogen to accumulate in the core at the expense of carbon and oxygen.
	These changes are subsequently transferred to the surface through rotational mixing and the removal of outer envelopes, as outlined by \citet{przybilla10} and \citet{maeder14}.
	Consequently, the new models (with reduced mass loss and mixing) exhibit smaller mass fractions of nitrogen at the end of the main sequence, while larger mass fractions of carbon and oxygen are observed.
	The less homogeneous structure is also expressed in lower fractions for the convective core mass (Fig.~\ref{coremass_z20}), because of the larger total mass.
	
	The differences between the new evolution models and the ones from \citet{yusof22} are more prominent as we escalate in the initial mass.
	For $M_\text{zams}=32\,M_\odot$ and above, the reduction of stellar mass at the end of the H-core burning is more moderate for the new models.
	Adopting updated mass-loss rate prescriptions, our model with an initial mass of $M_\text{zams}=32\,M_\odot$ loses only 1 $M_\odot$ during the main sequence, compared to  $\approx 5 M_\odot$ predicted by the old model.
	Similarly, our model with $M_\text{zams}=60$ loses only $\approx  5 M_\odot$ during the main sequence, compared to $\approx  25 M_\odot$ predicted by the old model.
	Hence, while \citet{yusof22} predict stars to lose up to half of their initial masses during the H-core burning, our new models predict a loss of only $\sim8\%$ at the most\footnote{This comparison is done, of course, considering the maximum mass $M_\text{zams}=60\,M_\odot$ in our analysis. More massive cases would develop WR winds during the main sequence, either from extreme removal of their outer envelopes or due to the proximity to the Eddington limit \citep[see][]{alex24b}, so we do not include them here.}.
	Such significant differences suggest substantial changes in subsequent evolutionary stages, even if other wind recipes remained unchanged, as will be discussed further.

\subsection{Evolution post main sequence stage}\label{evolutionHeC}
\subsubsection{H-shell burning stage}\label{hg}
        \begin{figure}[t!]
                \centering
                \includegraphics[height=0.9\linewidth]{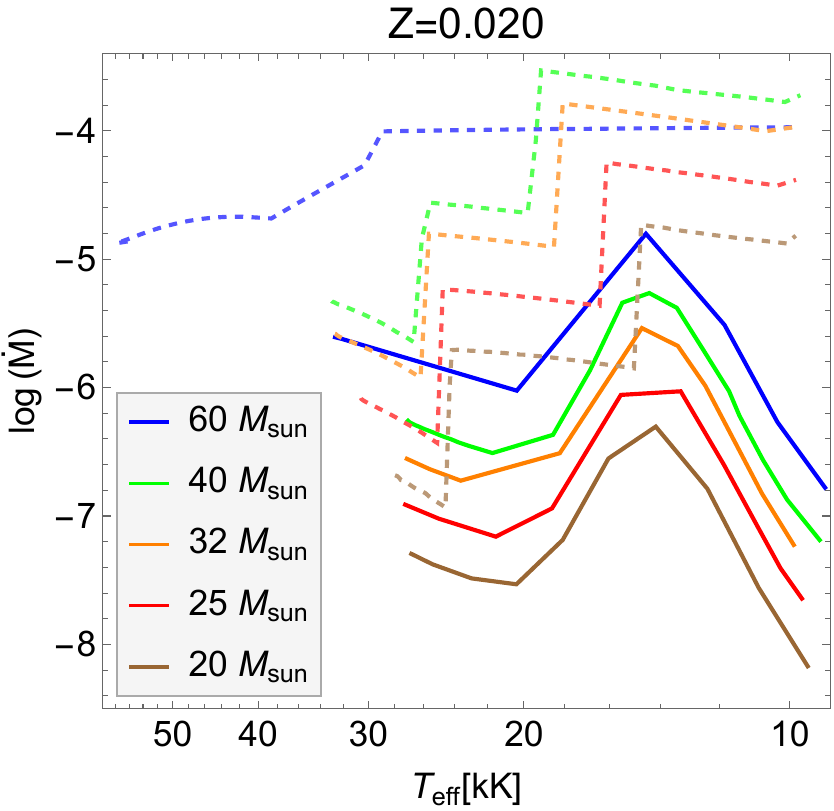}
                \caption{\small{Mass-loss rates after the H-depletion, as a function of the effective temperature, for the evolution tracks from \citet[dashed lines]{yusof22} and the tracks from this work (solid lines).
		Solid and dashed lines represent new and old models respectively, same as for Fig.~\ref{HRD_z20}.}}
                \label{midjump_vs_nojump}
        \end{figure}
	\begin{table}[t!]
		\centering
		\caption{Time interval and mass lost during the crossing of the Hertzsprung gap, according to our models.}
		\begin{tabular}{cc|cc}
			\hline
			\hline
			$M_\text{zams}$ & & $\Delta t$ & $\Delta M_*$\\
			$[M_\odot]$ & & [kyr] & $[M_\odot]$\\
			\hline
			$60$ & Y22 & 3.872 & 0.166\\
			$60$ & this work & 2.642 & 0.008\\
			\\
			$40$ & Y22 & 5.571 & 0.546\\
			$40$ & this work & 3.960 & 0.003\\
			\\
			$32$ & Y22 & 6.491 & 0.256\\
			$32$ & this work & 6.063 & 0.004\\
			\\
			$25$ & Y22 & 10.415 & 0.166\\
			$25$ & this work & 8.976 & 0.0019\\
			\\
			$20$ & Y22 & 16.696 & 0.090\\
			$20$ & this work & 12.947 & 0.0015\\
			\hline
		\end{tabular}
		\label{table_dMdt}
	\end{table}
	After the hydrogen depletion at the core, the H-shell burning stage begins, contracting the (now) He-core and expanding the outer layers \citep{groh14}.
	For this reason, all of our models quickly expand radially crossing the HR diagram horizontally at almost constant luminosity, the so-called Hertzsprung gap \citep[HG,][]{hurley00}, reaching temperatures below 10 kK at the end of the HG.
	Crossing $T_\text{eff}=10$ kK implies a change in the mass-loss rate recipe from \citetalias{krticka24} to \citet{dejager88}, but the rapid expansion just culminates when the He-core reaches temperatures high enough to start the He-burning process (He-ignition), which occurs at the red region of the HR diagram as seen in Fig.~\ref{HRD_z20}.
	
	During this expansion the stars experience a peak in the mass-loss rate at $T_\text{eff}\sim15$ kK according to the \citetalias{krticka24} formula \eqref{mdot_K24}, and thus the removed mass is lower than the predicted by previous models due to the bi-stability jump.
	This is explicitly evidenced in Fig.~\ref{midjump_vs_nojump}, where not only the overall mass loss is lower for new models, but also the jumps predicted by \citetalias{vink01} do not longer exist.
	Differences in the mass loss crossing the HG can reach up to two orders of magnitude, with the new values for $\dot M\sim10^{-7}-10^{-6}$ $M_\odot$~yr$^{-1}$ more in agreement with the m-CAK $\delta$-slow wind solutions for the regime of B-supergiants \citep{araya21,venero24}.
	Besides, the radial expansion across the Hertzsprung gap is faster because of the choice of the Ledoux criterion instead of Schwarzschild's: the (molecular weight) $\mu$-gradient at the border of the core prevents convection in the deep layers, weakening the support of the star that hence crosses rapidly the Hertzsprung gap \citep{sibony23}.
	In the case of the 20\,{\msol,} this happens too quickly (relatively to the Kelvin-Helmholtz timescale) for the thermal equilibrium to happen, so its radiation goes into expanding the envelope, which explains the drop in luminosity.
	Table~\ref{table_dMdt} shows the time required for the star to evolve from the end of the main sequence to the end of the HG: the lower mass loss combined with the shorter cross means that new models lose an amount of mass almost negligible when they cross the HG.
	Consequently, the envelope removal is almost negligible, so we can assume that the chemical composition of the stars is almost the same as the one tabulated for the end of the H-core burning.
	
\subsubsection{He-core and C-core burning stages}\label{He-burning}
\begin{figure*}[t!]
		\centering
		\includegraphics[width=0.22\linewidth,valign=c]{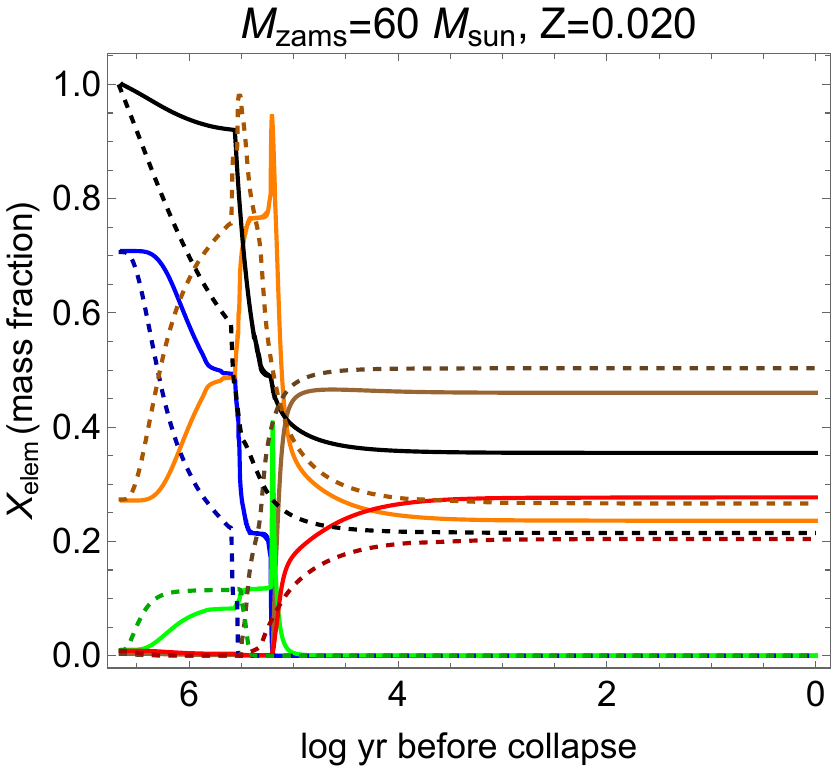}
		\hspace{1mm}
		\includegraphics[width=0.22\linewidth,valign=c]{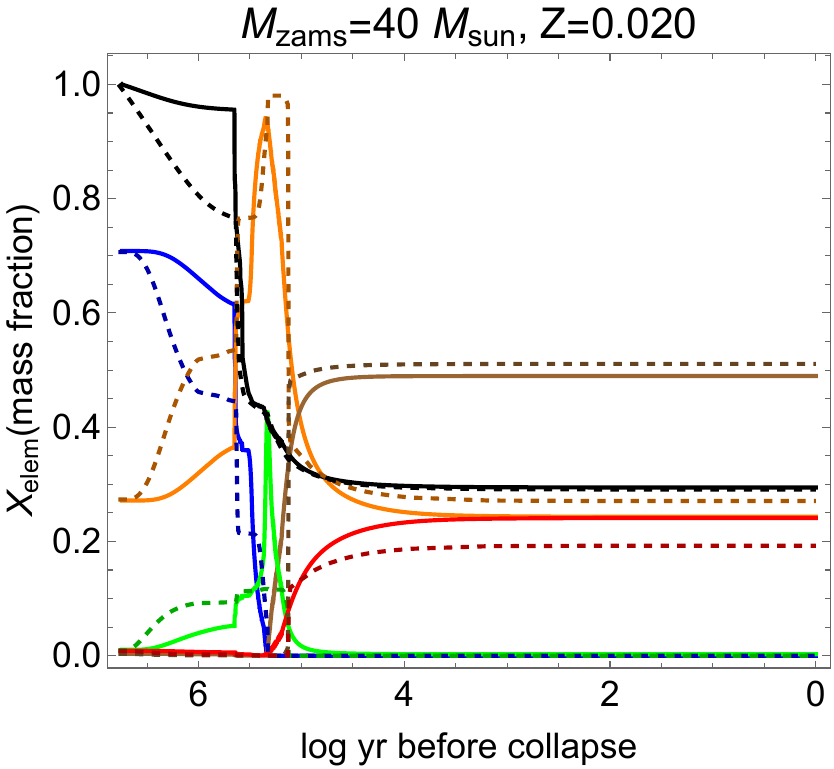}
		\hspace{1mm}
		\includegraphics[width=0.22\linewidth,valign=c]{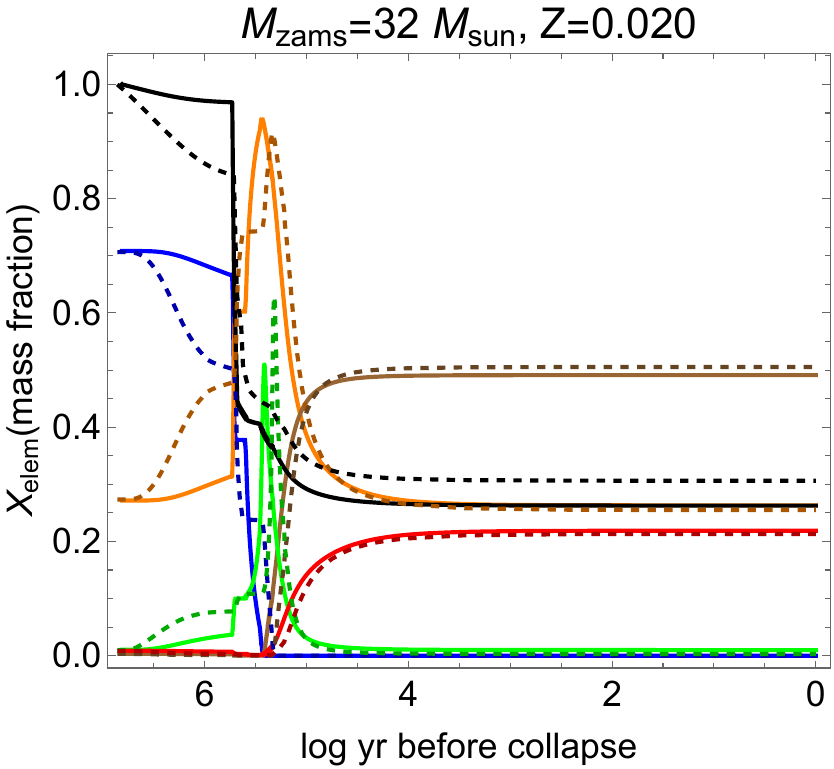}
		\hspace{1mm}
		\includegraphics[width=0.22\linewidth,valign=c]{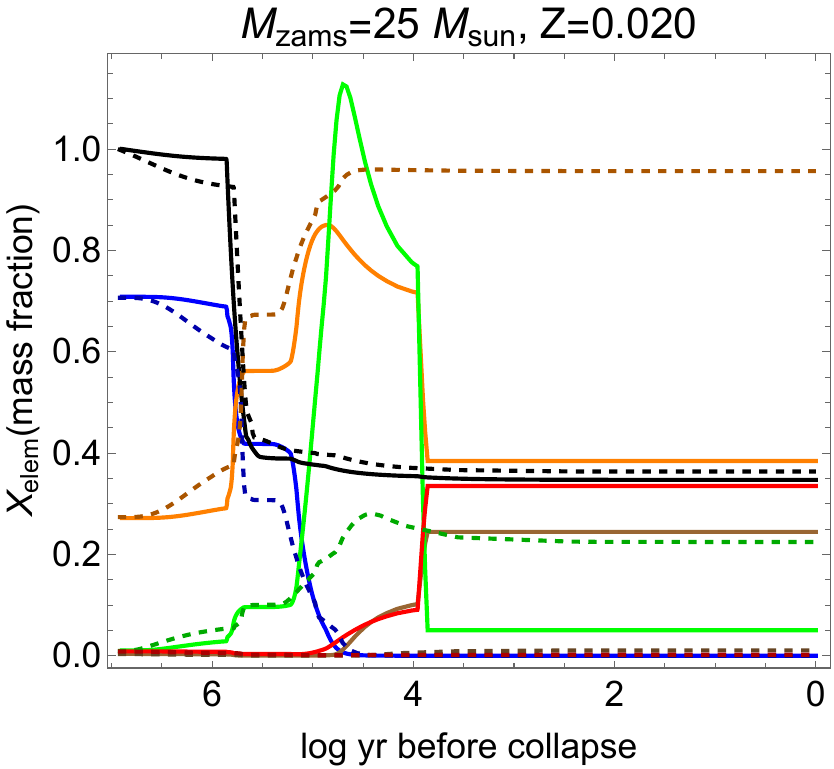}
		\includegraphics[width=0.07\linewidth,valign=c]{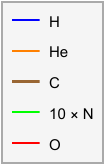}
		\caption{\small{Evolution of the surface abundances for hydrogen, helium, carbon, nitrogen, and oxygen, for the models reaching the WR phase.
		Black lines represent the total mass fraction, $M_*/M_\text{zams}$.
		The nitrogen abundance is multiplied by 10, for illustrative purposes.}}
		\label{reverse_abundances}
	\end{figure*}

	The higher $\dot M$ from \citet{yang23} makes our model of $M_\text{zams}=40\,M_\odot$ to experience a strong mass removal of the order of $\log\dot M_\text{Y23}\simeq-1.8$ when it crosses the threshold of $T_\text{eff}=5$ kK, with a subsequent drift back to $T_\text{eff}>5$ kK, thus entering in a loop between both wind regimes \citetalias{dejager88} and \citetalias{yang23}.
	This is a numerical artefact, but it can be physically interpreted as our $40\,M_\odot$ being subject to strong and variable outflows around $\sim 5$ kK, which prevents the star from becoming a RSG.
	Hence, the higher $\dot M$ from \citet{yang23} for $T_\text{eff}<5$ kK helps to preserve the upper Humphreys-Davidson limit \citep{humphreys16} on $\log (L/L_\odot)\simeq5.5$ from \citet{davies18} and \citet{mcdonald22}, besides also reducing the total time spent in the RSG phase \citep{zapartas24}.
	Indeed, the higher $\dot M_\text{Y23}$ produces the biggest mass removal during the entire evolution of our stars (as shown by the black lines in Fig.~\ref{reverse_abundances}).
	Additionally, the more efficient peeling-off effect during the RSG phase reduces the minimum initial mass required for a star to become a Wolf-Rayet.
	RSGs evolve blueward in the HR diagram when the mass of their convective cores reaches $60\%$ of the total mass \citep{meynet15}.
	Even though new models start their He-ignition with lower $M_\text{core}/M_*$ ratios (left panel of Fig.~\ref{coremass_z20}), the extra removal of outer envelopes compensates this condition and thus all stars with $M_\text{zams}\ge25\,M_\odot$ will continue evolving into WR stars.
	The exception is our $20\,M_\odot$ model, which reaches the RSG phase with a luminosity $\log (L/L_\odot)<5.0$ \citep[in contrast with the $\simeq5.2$ from][]{yusof22}, thus resulting in a lower $\dot M_\text{RSG}$ and in a poorer chemical enrichment at the stellar surface, at the end of the He-core and C-core burning stages.

	The rest of our models continue evolving towards the Wolf-Rayet phases.
	For the stars with 25 and 32 solar masses, new models predict lower final masses at the end of their lives (8.68 and 8.41 $M_\odot$ respectively), which are even lower than the final mass predicted for $M_\text{zams}=20\,M_\odot$.
	After the RSG phase, which corresponds to the abrupt drops in mass fraction and hydrogen, new models with $25$ and $32\,M_\odot$ end up with lower masses (compared with old models), whereas the opposite happens for models with $40$ and $60\,M_\odot$.
	And yet all models leave the RSG phase with larger H/He ratio, as a consequence of the lower mixing during the earlier evolutionary stages.
	According to \citet{nugis00} and \citet{grafener08} recipes, $\dot M_\text{WR}$ is more moderate for stars with larger H/He, and thus WR stars will initially lose less mass and envelope.
	However, this initially lower $\dot M_\text{WR}$ creates a larger inhomogeneity in the inner stellar structure, making evolution tracks to move more redwards during their WR lifetime, as noted in the HR diagram, and thus experiencing inflation episodes due to supra-Eddington layers \citep{ekstrom12}.
	As a result, the envelope removal of these WR stars is speeded up as they continue evolving.
	The case of $M_\text{zams}=25\,M_\odot$ is particularly noteworthy, because it makes the new track to continue evolving towards the WC phase (where carbon and oxygen become $\sim70\%$ of the surface composition), in contrast with the old model which predicts a final state as an almost pure He-star ($\approx96\%$ of its surface composition).

	For the most massive cases, this effect is less relevant.
	The model with 32 $M_\odot$ ends up with less mass but with a chemical composition at the end of the C-core burning stage comparable to the one predicted by \citet{yusof22}.
	For 40 $M_\odot$, $\dot M_\text{Y23}$ prevented the model from becoming an RSG star, but still resulted in a significant mass removal, compensating for initial deviations during the OB-type mass-loss phase, and explaining why the final mass and final chemical compositions are quite similar for both old and new models.
	For 60 $M_\odot$, the deviations during the main sequence were substantial enough to persist through subsequent stages, particularly since it never became an RSG, resulting in a final mass predicted by the new model ($21.3\,M_\odot$) that is approximately 65\% higher than that of the old model.

\section{Stellar properties at the Galactic Centre}\label{stellarproperties}
	We now proceed to evaluate the current evolutionary status of the massive stars at the Galactic Centre, according to our new evolution tracks.
        We are particularly interested in the chemical surface abundance of these stars, given their influence on the accretion on to the central super-massive black hole \citep[e.g.,][]{calderon25}.
        Despite the known  kinematics of hundreds of massive stars in the central parsec of the Milky Way \citep{vonfellenberg22}, the only stars with measured stellar properties are the sample of WR stars from \citet{martins07} and the so-called `S-stars' from \citet{habibi17}.
        These S-stars, however, are spectroscopically B-dwarfs (i.e., early main sequence stars without any wind) with initial masses between $\sim7-15$ $M_\odot$, whose evolutionary status is barely affected by updated wind mass-loss rate prescriptions.
        Therefore, our observational analysis is focused on the sample of WR stars.

\subsection{Theoretical predictions for observed WR stars}\label{wrstars}
        \begin{figure*}[t!]
                \centering
                \includegraphics[height=0.36\linewidth]{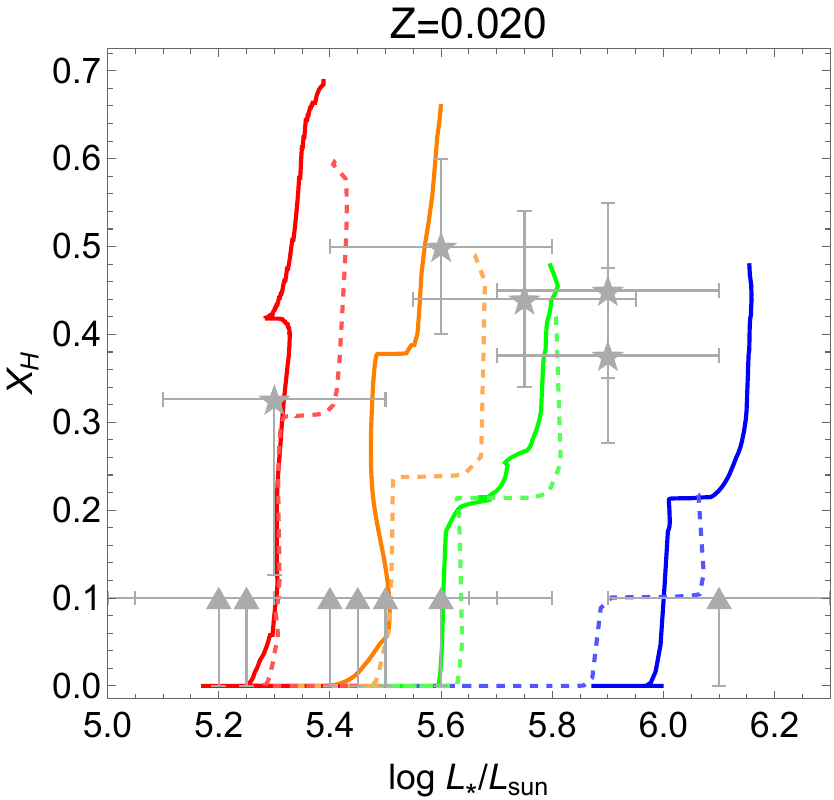}
                \hspace{1cm}
                \includegraphics[height=0.36\linewidth]{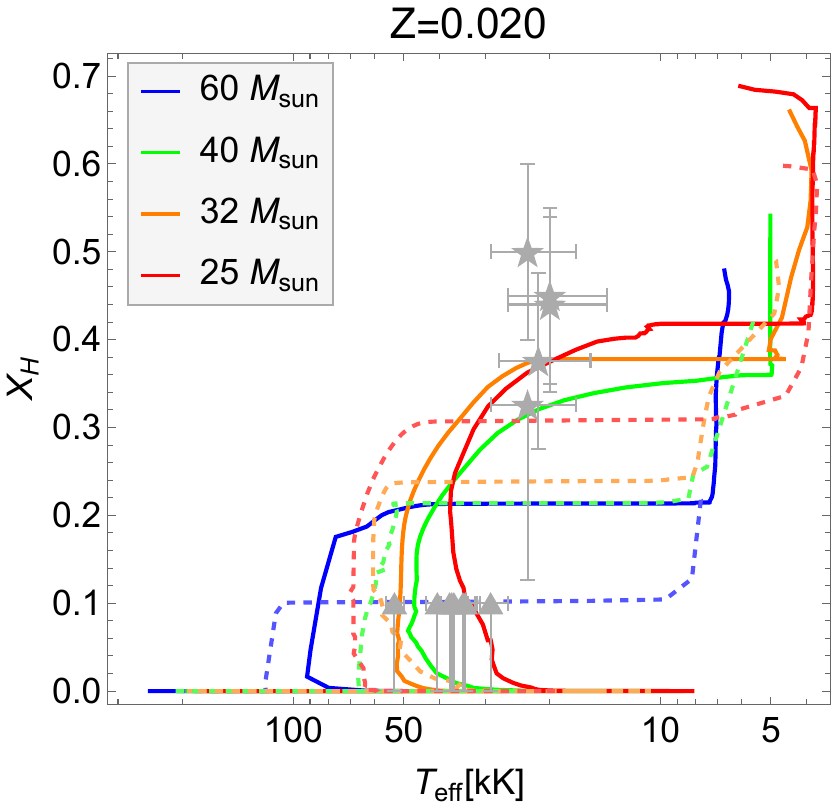}
                \caption{\small{Left panel: hydrogen mass fraction at the stellar surface as a function of the stellar luminosity, from the H-core ignition to the C-core depletion.
                Right panel: hydrogen mass fraction at the stellar surface as a function of the stellar temperature, from He-ignition to the C-core depletion.
                Grey stars and triangles represent the observational values calculated by \citet{martins07}, for Ofpe/WN9 and WN stars respectively.
                Solid and dashed lines represent new and old models respectively, same as for Fig.~\ref{HRD_z20}.}}
                \label{ofpe_martins}
        \end{figure*}
        \begin{figure*}[t!]
                \centering
                \includegraphics[height=0.36\linewidth]{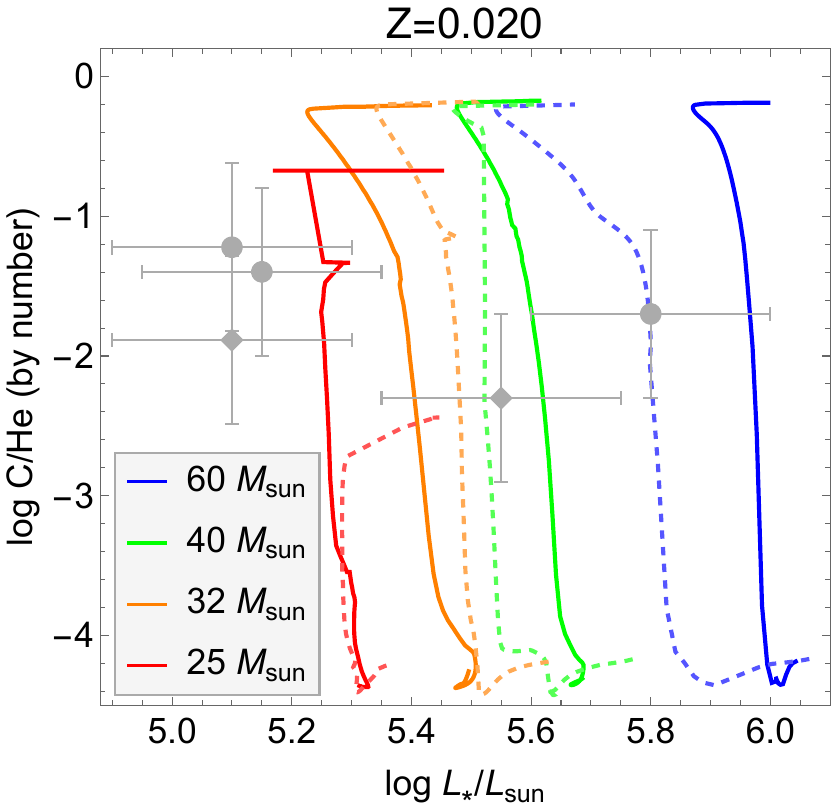}
                \hspace{1cm}
                \includegraphics[height=0.36\linewidth]{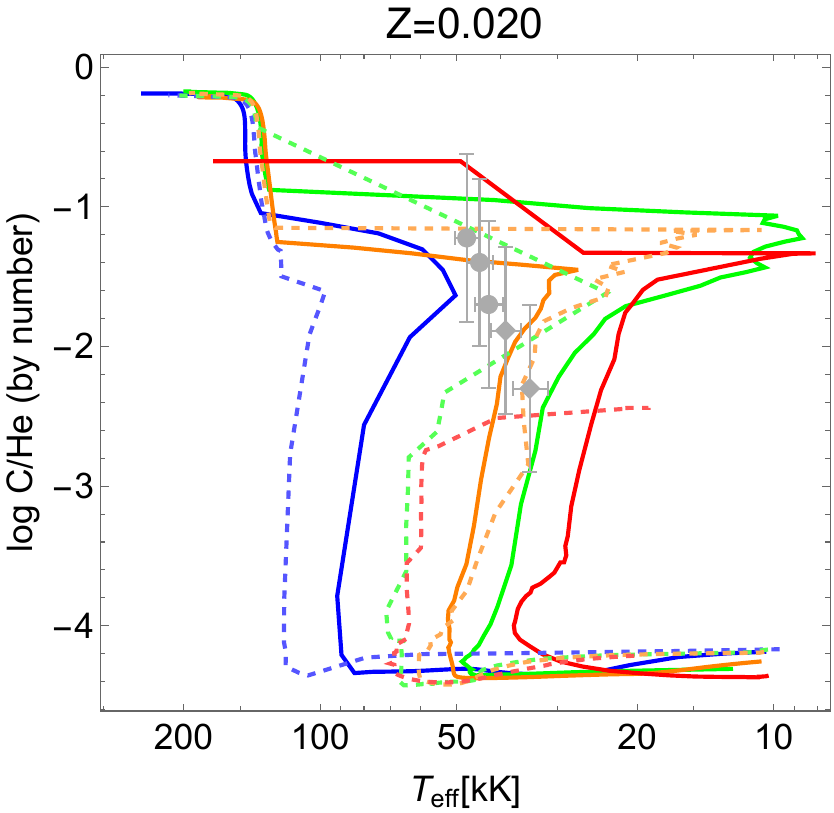}
                \caption{\small{Left panel: carbon-to-helium ratio (by number) at the stellar surface as a function of the stellar luminosity, from post YSG/RSG phase to the C-depletion.
                Right panel: carbon-to-helium ratio (by number) at the stellar surface as a function of the stellar temperature, from post YSG/RSG phase to the C-depletion.
                Grey diamonds and circles represent the observational values calculated by \citet{martins07}, for WN/C and WC stars respectively.
                Solid and dashed lines represent new and old models respectively, same as for Fig.~\ref{HRD_z20}.}}
                \label{wnc_martins}
	\end{figure*}
	
	For the case of WR stars at the Galactic Centre, the known sample includes spectral types Ofpe/WN9, WN (subtypes 5 to 8), WN/C, and WC (subtypes 8 and 9).
	Their determined temperatures, luminosities, and abundances are given by \citet[see their Table~2]{martins07}.
	
\subsubsection{Ofpe/WN9 and WN stars}\label{ofpewnstars}
	The left panel of Fig.~\ref{ofpe_martins} shows the hydrogen fraction as a function of luminosity, starting from the He-ignition, for both old and new evolution tracks.
	This figure is analogous to Fig.~21 by \citet{martins07}, where the used tracks were from \citet{meynet05}.
	From here, it was deduced that Ofpe/WN9 are immediate precursors of WR stars.
	However, because the post-MS stellar evolution for stars in the range of 25 to 60 $M_\odot$ is at almost constant luminosity, the plane $X_\text{surf}-L_*$ is not enough to provide insights on the evolution status of these stars.
	For that reason, we add a second panel on the right showing the hydrogen mass fraction but as a function of the stellar temperature \citep[the radius rescaling is taken from][]{meynet05}, also from the start of the He-core burning.
	Given that all our models start their He-ignition with $T_\text{eff}<10$ kK (Fig.~\ref{HRD_z20}), they begin being YSG/RSG stars losing surface hydrogen at  almost constant temperature before continuing evolving bluewards post-YSG/RSG.
	New models predict post-YSG/RSG evolution to be H-richer than their counterparts predicted by \citet{yusof22}, making them also redder.
	Hence, we can see that the hydrogen mass fractions predicted by our evolution models, for both post-RSG and post-YSG phases, do a good match with the high $X_\text{H}$ exhibited by these Ofpe/WN9 stars.
	This, notwithstanding the possibility of some of these Ofpe/WN9 being in a quiescent LBV state   \citep{crowther95}, in particular the stars with the largest hydrogen content.
	
	Certainly all the evidence supports that Ofpe/WN9 are precursors of late-type WN stars, but this does not necessarily mean that WN8 will be the immediate stage after.
	Figure~\ref{ofpe_martins} also shows that, with the exception of the 60 $M_\odot$ model, the new tracks for post Ofpe/WN9 phase start decreasing their temperatures as the surface hydrogen goes below $\sim20\%$.
	If we consider $X_\text{H}=0.2$ to be the threshold between Ofpe/WN9 and WNL stars \citep[a reasonable assumption, given the observed abundances of the][sample]{martins07}, we can suggest that the temporal sequence for the initial mass range $\sim25-40$ $M_\odot$ is not descendent from WN8 to earlier WNL (subtypes 7 and 6) but the opposite (Ofpe/WN9$\rightarrow$WN6$\rightarrow$WN7$\rightarrow$WN8$\rightarrow$...), or even non monotonic.
	Naturally this statement is not conclusive given that the actual distinctions between the WN subtypes are not based on evolutionary properties but on the properties of the He and N lines \citep{smith96,crowther07}.
	However, the surface nitrogen abundance actually increases from the beginning of the WR phase (Fig.~\ref{reverse_abundances}) whereas observed WN8 and WN8/WC9 stars contain more nitrogen than their earlier siblings WN 5/6 and WN 7 \citep[][Table~2]{martins07}, thus supporting our assessment.
	This means, WN8 stars at the GC are closer to the other subtypes WN5-7 rather than to Ofpe/WN9, this is the reason why they should be included in the same WNL subcategory for posteriori wind colliding models \citep[compare with the input of sub-groups of WR stars adopted by][]{calderon25}.

\subsubsection{WN/C and WC stars}
	WN/C stars are WR stars which exhibit an equilibrium between nitrogen and carbon at their surfaces (C/N ratio of the order of 1).
	This spectral type is expected to be found for stars born with masses between $\sim30$ and 60 $M_\odot$, massive enough to show He-burning nuclear waste (carbon) before their core collapse, but not so massive to not be quickly depleted of all its surface hydrogen before becoming WC \citep{meynet03}.
	When the WR phase starts, the stellar surface is very C-poor because the CNO cycle converts carbon into nitrogen, but it becomes C-rich as the products of the He-burning (carbon and oxygen) start appearing in the outer envelopes.
	Nitrogen abundance also increases during the WR phase, as shown by Fig.~\ref{reverse_abundances}, but at a smaller rate than carbon, reason why this ratio C/N grows up.

	Such evolutionary diagnostics are based on solar metallicity (assuming $Z_\odot=0.02$), but still fits quite well with our set of new evolution models.
	Figure~\ref{wnc_martins} shows the evolution of the C/He ratio post-YSG/RSG as a function of the stellar luminosity (left panel) and temperature (right panel), for both new and old tracks.
	The surface carbon abundance increases as the WR stars slightly decrease their temperatures, with the WN/C stars from \citet{martins07} located prior to the end of this redwards drift.
	As for Martins' WC stars, the C/He ratio coincides with the rapid jump in temperature in the new models which overcomes with the removal of the last traces of hydrogen (i.e., with the end of any WN signature).
	Still, this behaviour of C/He vs $T_\text{eff}$ is found for both old and new evolution tracks.
	The differences come when we analyse the C/He ratio agains the luminosity.
	Given that our 25 $M_\odot$ model continues evolving towards the WC phase, it matches better the less bright WN/C and WC stars analysed by \citet{martins07}, without the need of invoking binary evolution.
	
	If the latest traces of hydrogen are still present in the outer envelope of our structure models, the calculated opacity for this outer envelope becomes uncertain and subsequently the radius can be numerically inflated, with the respective drop in the temperature.
	This opacity effect is an inherent condition produced by the mismatch between (hydrostatic) evolution models and (expanding) atmospheric models.
	Even though `deep' expanding atmospheric models can correct the expected temperatures predicted in \textsc{Genec} models subject to envelope inflation, it requires a considerable computational effort and it has been implemented only for the main sequence stage of very massive stars \citep{josiek25}.
	However, this opacity effect represents a minimal fraction of the WR lifetime, and does not interfere with the chemical composition nor temperatures measured for the WN/C and WC stars in the sample. 
	Therefore, despite this issue and the associated error bars in the infrared spectra, new evolution tracks at super solar metallicity can adequately reproduce the observational diagnostics for the massive stars of the Galactic Centre.

\subsection{Evolutionary phases}
	\begin{figure}[t!]
		\centering
		\includegraphics[width=0.9\linewidth]{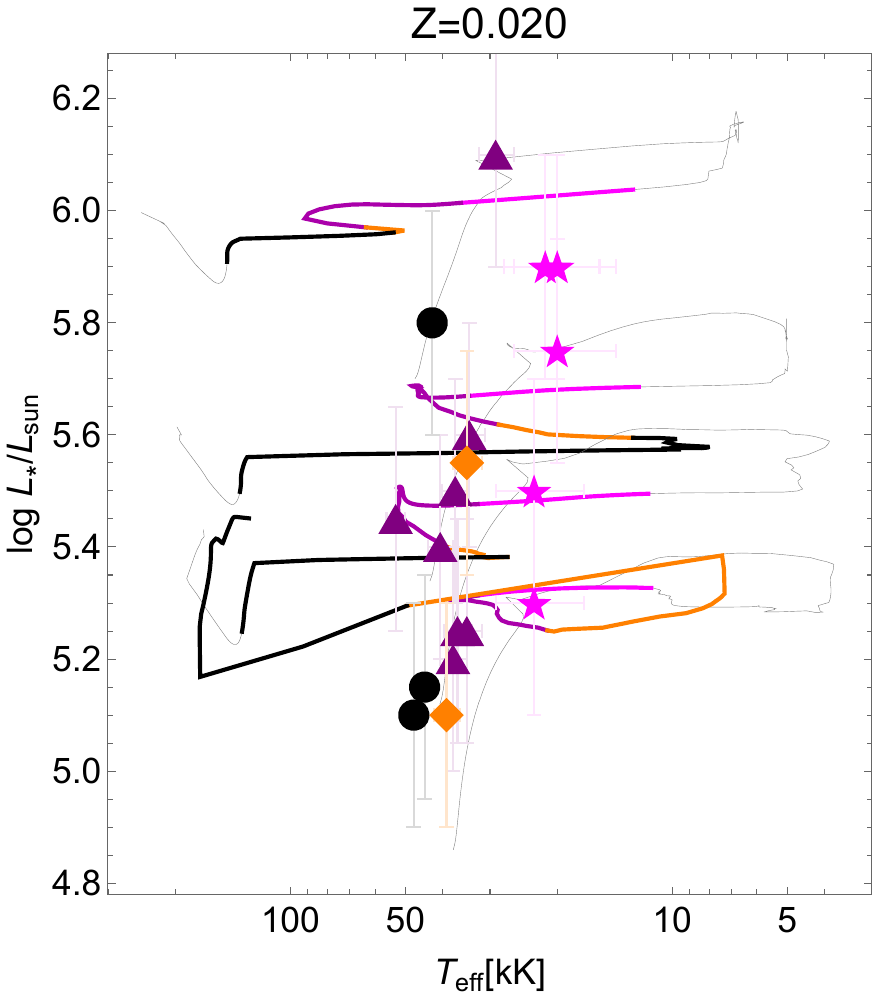}
		\caption{\small{HR diagram showing the position of the spectroscopic phases Ofpe/WN9 (magenta), WN (purple), WN/C (orange) and WC (black).
		Observed stars from \citet{martins07} are shown, with the same symbols as Fig.~\ref{ofpe_martins} and Fig.~\ref{wnc_martins}.}}
		\label{HRD_z20_wstars}
        \end{figure}

	Following \citet{martinet23}, we proceed to define the WR subtypes as evolutionary stages.
	Since we cannot use spectroscopic criteria for these categories (because we are not providing synthetic spectra), we use as criteria some abundances and the temperature calculated by our models.
	Hence
	\begin{itemize}
		\item \text{Ofpe/WN9}: WR star with $X_\text{H}\ge0.2$ and $T_*\le30$ kK.
		\item \text{eWNL}: WR star with $X_\text{H}\ge10^{-5}$.
		\item \text{eWNE}: WR star and $X_\text{H}<10^{-5}$ and $X_\text{N}>X_\text{C}$.
		\item WN/C: WR star with $-0.5<\log(X_\text{C}/X_\text{N})<0.5$.
		\item WC: WR star with $\log T_*<5.20$ and $(X_\text{C}+X_\text{O})/X_\text{He}<1.6$.
	\end{itemize}
	
	The threshold for Ofpe/WN9 stars come from the analysis of Sec.~\ref{ofpewnstars} and from the parameters calculated by \citet{crowther95}.
	The boundary $X_\text{H}=10^{-5}$ for eWNL and eWNE \citep[where the `e' stands for `evolutionary', following][]{foellmi03} comes from \citet{georgy12}.
	The boundaries for the WN/C phase are in rule with the C/N values reported by \citet{crowther95b}.
	The separation between WC and WO comes from the C, O, and He abundances measured by \citet{aadland22}.
	
\begin{figure}[t!]
		\centering
		\includegraphics[width=0.9\linewidth]{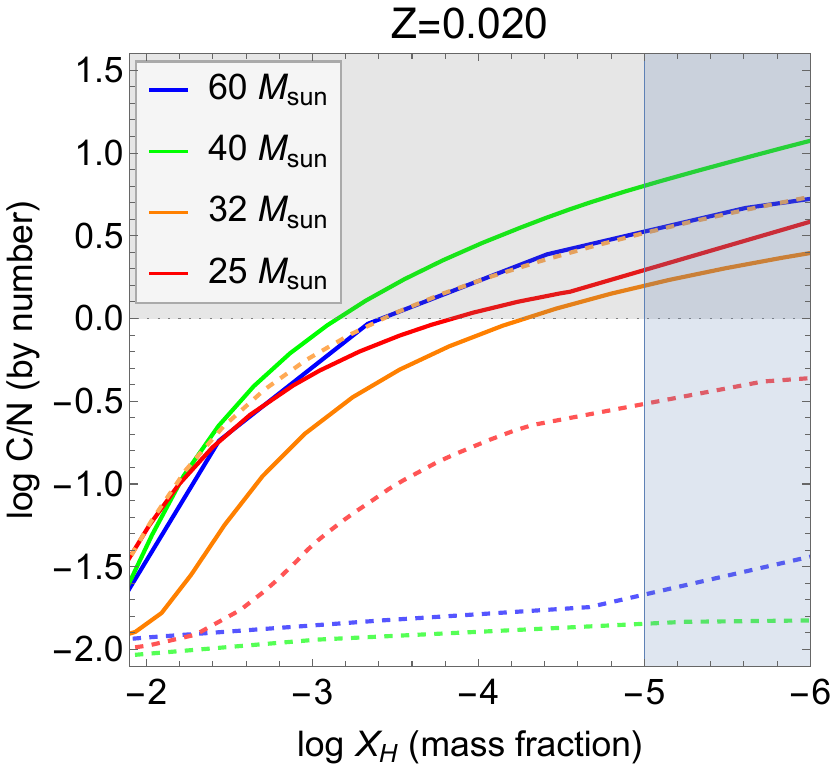}
		\caption{\small{Evolution of the C/N ratio in evolution models, as the stars are losing their last $1\%$ of hydrogen in the outer layer.
		Blue-shaded region represents $X_\text{H}\le10^{-5}$ and $X_\text{C}\ge X_\text{N}$, whereas grey-shaded region $X_\text{H}\ge10^{-5}$ and $X_\text{C}\le X_\text{N}$.
		Solid and dashed lines represent new and old models respectively, same as for Fig.~\ref{HRD_z20}.}}
		\label{CNratio_vs_Xs}
        \end{figure}

	We highlight the segment of each one of the Wolf-Rayet phases (Ofpe/WN9, WNL, WN/C and WC) across the HR diagram in Fig.~\ref{HRD_z20_wstars}, together with the GC stars from \citet{martins07}.
	In general we notice a consistency between observed and evolutionary spectral type, with some minor discrepancies due to the lack of unified criteria.
	For example, our C/N abundances for the WN/C phase differ from the boundaries used by \citet{georgy12}, where WN/C stars are additionally expected to be H-poor.
	In our case, given the moderate chemical evolution consequence of lower winds, we still find traces of hydrogen at the surface of WN/C stars.
	This is observed in Fig.~\ref{CNratio_vs_Xs}, where new evolution models predict stars to become carbon-richer prior to the total depletion of hydrogen from the stellar surface.
	In contrast, old models predict stars with $X_\text{surf}\le10^{-5}$ and $X_\text{N}>X_\text{C}$, thus being in the category of eWNE stars.
	Thus, all our evolution models predict that (H-rich) WNL stars evolve into WN/C without passing through eWNE phase.
	This is consistent with the lack of early type WN stars in the sample of \citet{martins07}, where the earliest WN star is 16SE2 with spectroscopic subtype WN5/6.
	Besides, as mentioned in Sec.~\ref{ofpewnstars}, the low nitrogen content of 16SE2 suggests that this WN5/6 might be located before the WN7-8 stars in the evolutionary sequence, strengthening the idea of the WNL phase being the direct precursor of WN/C for the mass and metallicity range  analysed in this work.
	Again, the aforementioned by-pass to the eWNE phase is based only on the abundance criteria coming from the evolution models instead of from spectroscopic features, so it cannot be taken as a full absence of WNE stars in the GC.

	Now we proceed to calculate the surface abundances for hydrogen, helium, carbon, nitrogen, and oxygen, for our evolution models during Ofpe/WN9, WNL, WN/C, and WC phases.
	Given that the individual surface abundances are still changing within the same spectral subtypes, we calculate the time-averaged abundances, defined as
	\begin{equation}
		\langle X_\text{elem}\rangle=\frac{1}{\Delta\tau_\text{phase}}\int_{t_\text{i}}^{t_\text{f}} X_\text{elem}(t)\,dt\;,
	\end{equation}
	where $\Delta\tau_\text{phase}=t_\text{f}-t_\text{i}$ ($t_\text{f}$ -- final time, $t_\text{i}$ -- initial time) is the lifetime of the respective subtype, and the corresponding uncertainties as the standard deviations,
	\begin{equation}
		\sigma(X_\text{elem})=\sqrt{\frac{1}{\Delta\tau_\text{phase}}\int_{t_\text{i}}^{t_\text{f}}[X_\text{elem}(t)-\langle X_\text{elem}\rangle]^2\,dt}\;.
	\end{equation}
	
	Results for the WR phases are shown in Table~\ref{table_abundances}, together with the lifetime of the spectroscopic phases and the average mass-loss rate (calculated as $\Delta M_{*,\text{lost}}/\Delta\tau_\text{phase}$).
	WR stars are less chemically evolved at their surface for lower initial masses.
	For Ofpe/WN9 phase, the model with $M_\text{zams}=25\,M_\odot$ presents the largest hydrogen mass fraction, given that it experienced less mass removal than their counterparts, which was also seen in the right panel of Fig.~\ref{ofpe_martins}.
	It also has the longest lifetime given that the mass loss during this Ofpe/WN9 is the most reduced.
	For the WN phase, the abundances have a larger standard deviation as a consequence of the quick changes in the chemical composition.
	For WN/C and WC stars, the abundances are mostly of the same orders of magnitude, with the notable exception of $M_\text{zams}=25\,M_\odot$, which is the only model reaching  the end of the He-core burning stage in the WN/C phase and then it becomes a WC star only at the very end prior to the end of C-core burning.

	\begin{table*}[t!]
		\centering
		\caption{Time-averaged surface abundances at the different WR spectroscopic phases of our evolution models.}
		\resizebox{\linewidth}{!}{
		\begin{tabular}{ccc|ccccc|c}
			\hline\hline
			Spectral phase & $M_\text{zams}$ & $\Delta\tau_\text{phase}$ & $\langle X_\text{H}\rangle$ & $\langle X_\text{He}\rangle$ & $\langle X_\text{C}\rangle$ & $\langle X_\text{N}\rangle$ & $\langle X_\text{O}\rangle$ & $\log\langle\dot M\rangle$\\
			& [$M_\odot$] & [yr] & \multicolumn{5}{c}{mass percentage} & [$M_\odot$ yr$^{-1}$]\\
			\hline
			\multicolumn{3}{c}{ZAMS, $Z=0.020$} & 70.83 & 27.16 & 0.354 & 0.103 & 0.858\\
			\hline
			Ofpe/WN9 & 60 & $8.61\times10^3$ & $21.35\pm0.01$ & $76.69\pm0.01$ & $0.013\pm0$ & $1.173\pm0$ & $0.090\pm0$ & $-3.91$\\
			& 40 & $1.60\times10^4$ & $31.96\pm2.11$ & $66.07\pm2.11$ & $0.010\pm0$ & $1.089\pm0.02$ & $0.191\pm0.02$ & $-4.61$\\
			& 32 & $4.44\times10^4$ & $37.27\pm1.02$ & $60.75\pm1.02$ & $0.009\pm0$ & $1.007\pm0.01$ & $0.286\pm0.01$ & $-5.00$\\
			& 25 & $9.37\times10^4$ & $39.91\pm2.38$ & $58.04\pm2.29$ & $0.008\pm0$ & $1.022\pm0.10$ & $0.325\pm0.01$ & $-5.80$\\
			\hline
			eWNL & 60 & $4.72\times10^4$ & $18.01\pm6.04$ & $79.78\pm5.39$ & $0.035\pm0.13$ & $1.365\pm0.59$ & $0.104\pm0.10$ & $-4.34$\\
			& 40 & $9.17\times10^4$ & $10.52\pm7.00$ & $86.88\pm6.31$ & $0.067\pm0.20$ & $1.695\pm0.87$ & $0.105\pm0.07$ & $-4.81$\\
			& 32 & $1.35\times10^5$ & $12.30\pm9.08$ & $84.73\pm8.07$ & $0.077\pm0.23$ & $1.975\pm1.21$ & $0.146\pm0.07$ & $-5.14$\\
			& 25 & $1.04\times10^5$ & $10.98\pm8.91$ & $80.29\pm4.88$ & $0.339\pm0.72$ & $6.062\pm3.29$ & $1.406\pm1.31$ & $-5.28$\\
			\hline
			WN/C & 60 & $4.14\times10^3$ & $0.048\pm0.06$ & $90.04\pm1.63$ & $4.176\pm1.74$ & $3.368\pm0.47$ & $1.327\pm0.34$ & $-3.85$\\
			& 40 & $1.66\times10^4$ & $0.076\pm0.07$ & $90.15\pm1.56$ & $4.391\pm1.81$ & $3.498\pm0.48$ & $0.741\pm0.21$ & $-4.44$\\
			& 32 & $3.06\times10^4$ & $0.009\pm0.01$ & $88.96\pm1.81$ & $5.101\pm2.04$ & $3.918\pm0.55$ & $0.780\pm0.25$ & $-4.80$\\
			& 25 & $3.02\times10^4$ & $0.046\pm0.05$ & $75.44\pm2.30$ & $7.053\pm1.98$ & $9.111\pm0.92$ & $6.997\pm1.22$ & $-5.16$\\
			\hline
			WC & 60 & $4.48\times10^4$ & $0\pm0$ & $56.10\pm15.10$ & $30.75\pm10.78$ & $0.661\pm0.60$ & $10.43\pm4.51$ & $-4.16$\\
			& 40 & $1.06\times10^5$ & $0\pm0$ & $62.40\pm16.24$ & $27.77\pm12.27$ & $0.959\pm0.75$ & $6.907\pm4.36$ & $-4.58$\\
			& 32 & $1.31\times10^5$ & $0\pm0$ & $59.74\pm15.32$ & $29.49\pm11.58$ & $1.23\pm0.83$ & $7.520\pm4.25$ & $-4.71$\\
			& 25 & $9.03\times10^3$ & $0\pm0.01$ & $36.28\pm9.41$ & $26.77\pm4.62$ & $1.03\pm1.69$ & $34.14\pm6.71$ & $-4.74$\\
			\hline
		\end{tabular}}
		\label{table_abundances}
	\end{table*}

	These tabulated abundances represent an important upgrade for the modelling of the Galactic centre environment around Sgr~A*.
	Recent works have adopted the chemical composition from a selection of observed WR stars, first compiled by \cite{russell17} and used in that study and by \citet{wang20,Balakrishnan24} to produce synthetic X-ray spectra from hydrodynamical simulations, and more recently by \citet{calderon25} and \citet{labaj25} as a simulation ingredient.
        Those stars were classified as either WC8-9, WN5-7, and WN8-9 and Ofpe/WN9 stars, with the last two subtypes treated as a single category, with $X_\text{H}=11.5$ and $X_\text{He}=82.5$.
	However, our evolution models suggest that WN8 may not be the direct descendent of Ofpe/WN9, and for that reason we consider they should not be combined.
	Indeed, Ofpe/WN9 stars are expected to have higher hydrogen abundances, in line with the spectral diagnostics performed by \citet{martins07}.
	Likewise, evolutionary WN5-8 stars as a whole category are now expected to contain a higher fraction of hydrogen due to their previous evolutionary stages, although with a large standard deviations.
	On the other side, the average abundances for WC are more in agreement with the data tabulated  by \citet{russell17}, with the exception of our $25\,M_\odot$ model for the reasons given in the previous paragraph.

	All in all, our new abundances lie in between those used previously in the literature for hydrodynamical simulations of the Galactic centre WR winds, namely either Solar with three times enhanced metallicity \citep[e.g.,][]{calderon20apj}, or the \citet{russell17} tabulated abundances \citep{calderon25}.  
	The former resulted in a higher rate of radiative cooling after the wind shock collisions, and the subsequent formation of clumps and even a cold disc that increased the accretion on to Sgr~A*, while in the latter those effects were suppressed. 
	Updated hydrodynamical models should be developed using the abundances obtained in this work, for which not only the spectral type of the stars can be used, but also estimates of their masses when available. 

\section{Summary and conclusions}\label{conclusions}
	We show a set of evolutionary tracks using the Geneva-evolution-code (\textsc{Genec}), for stars with initial masses from $20$ to $60\,M_\odot$, with new mass-loss rate recipes for the winds of OB-type stars \citep{krticka24} and RSGs \citep{yang23}.
	Because of the lower $\dot M$ during the main sequence stage and the Hertzsprung gap, stars retain more than $\approx92\%$ of their initial masses prior to the stronger dust-driven mass loss at the red part of the HRD, where the star finally losses most of its envelopes.
	
	These changes in the mass-loss history lead into Wolf-Rayet stars with a less evolved chemical evolution (and subsequently 'cooler') in comparison with previous models \citep{yusof22}.
	Post Ofpe/WN9 spectral phase, stars stay as late-type WN to become WN/C, by-passing the early-type WN spectral phase.
	This scenario agrees with the observed population of WR in the GC \citep{martins07}, where the late-type clearly dominates for all WN, WN/C and WC subtypes.
	Based on this scenario, we tabulate the respective mean surface abundances and lifetimes of each one of the WR spectroscopic phases, to provide a more robust framework for the study of the colliding winds from the population of WR stars around the supermassive black-hole Sgr~A* at the centre of the Milky Way \citep{calderon25}.
	
	The results presented in this work represent the state-of-the-art in terms of stellar wind theory, stellar evolution at high metallicity environments, and observations of the Galactic Centre.
	Considering that $Z=0.020$ represents a lower threshold for the probable metallicity of the centre of the Milky Way, further studies need to be performed with a wind theory capable of providing adequate values of mass loss at higher $Z$.
	Likewise, new spectroscopic measurements are needed for the population of WR stars in the GC in order to decrease the uncertainties in the currently known abundances as well as on their actual wind properties ($\dot M$ and $\varv_\infty$).
	Nevertheless, the new evolutionary tracks introduced in this work as well as the surface abundances predicted for WR stages represent an important upgrade in our understanding on the physical conditions in the Galactic Centre, and they will be the initial starting point for future population synthesis studies for colliding winds.

\begin{acknowledgements}
	We acknowledge useful discussions with Diego Calderón, Chris Russell, Wasif Shaqil, and Sebastiano von Fellenberg.
        ACGM thanks the support from project 10108195 MERIT (MSCA-COFUND Horizon Europe).
        ACGM and JC acknowledge the support from the Max Planck Society through a "Partner Group" grant.
        JC acknowledges financial support from ANID -- FONDECYT Regular 1251444, and Millennium Science Initiative Program NCN$2023\_002$.
        BK and JK acknowledge support from the Grant Agency of the Czech Republic (GAČR 25-15910S).
        SE acknowledges support from the Swiss National Science Foundation (SNSF), grant number 212143.
        The Astronomical Institute of the Czech Academy of Sciences in Ond\v rejov is supported by the project RVO: 67985815.
\end{acknowledgements}
\bibliography{gc_paper.bib} 
\bibliographystyle{aa} 

\
\begin{appendix}
\onecolumn
\section{Mass-loss rate recipes}\label{mdot_formulae}
	For the winds of OB-type stars ($T_\text{eff}\ge10$ kK) we use the formula from \citet[Eq.~3 with coefficients from Table~3]{krticka24}
	\begin{align}\label{mdot_K24}
		\log\dot M_\text{K24}=&-13.82+0.358\log\left(\frac{Z}{Z_\odot}\right)+\log\left(\frac{L_*}{10^6L_\odot}\right)\times\left[1.52-0.11\log\left(\frac{Z}{Z_\odot}\right)\right]\nonumber\\
		&+13.82\log\Biggl\{\left[1.0+0.73\log\left(\frac{Z}{Z_\odot}\right)\right]\exp\left[-\frac{(T_\text{eff}/\text{kK}-14.16)^2}{3.58^2}\right]+3.84\exp\left[-\frac{(T_\text{eff}/\text{kK}-37.9)^2}{56.5^2}\right]\Biggr\}\;.
	\end{align}
	
	For the mass-loss rate of stars in the range of temperatures between 5 and 10 kK we use the linearised formula of \citet{dejager88}
	\begin{equation}
		\log\dot M_\text{dJ88}=-8.158+1.769\log\left(\frac{L_*}{L_\odot}\right)-1.676\log T_\text{eff}\;.
	\end{equation}
	
	For the mass-loss rate of red supergiants ($T_\text{eff}\le5$ kK) we use the formula of \citet{yang23}
	\begin{equation}
		\log\dot M_\text{Y23}=0.45\left[\log\left(\frac{L_*}{L_\odot}\right)\right]^3-5.26\left[\log\left(\frac{L_*}{L_\odot}\right)\right]^2+20.93\log\left(\frac{L_*}{L_\odot}\right)-34.56\;.
	\end{equation}
	
	For the wind of Wolf-Rayet stars with $T_*\le70$ kK we use \citet{grafener08}
	\begin{equation}
		\log\dot M_\text{GH08}=10.046+\beta\log(\Gamma_\text{e}-\Gamma_0)-3.5\log(T_*/kK)+0.42\log\left(\frac{L_*}{L_\odot}\right)-0.45 X_\text{H}\:,
	\end{equation}
	with
	\begin{equation}
		\beta(Z)=1.727+0.25\log\left(\frac{Z}{Z_\odot}\right)\;,\nonumber
	\end{equation}
	\begin{equation}
		\Gamma_0(Z)=0.325-0.301\log\left(\frac{Z}{Z_\odot}\right)-0.045\left[\log\left(\frac{Z}{Z_\odot}\right)\right]^2\;.\nonumber
	\end{equation}
	
	Finally, for Wolf-Rayet stars out of the range covered \textbf{by} \citet{grafener08}, we use the formulae from \citet{nugis00} for WN and WC/WO phases
	\begin{equation}
		\log\dot M_\text{NL00,WN}=-13.6+1.63\log\left(\frac{L_*}{L_\odot}\right)+2.22\log Y_\text{He}\;,
	\end{equation}
	\begin{equation}
		\log\dot M_\text{NL00,WC/WO}=-8.30+0.84\log\left(\frac{L_*}{L_\odot}\right)+2.04\log Y_\text{He}+1.04\log\left(\frac{Z}{Z_\odot}\right)\;.
	\end{equation}
	
\end{appendix}

\end{document}